\documentclass[final, journal, twocolumn]{IEEEtran}
\usepackage[utf8]{inputenc}
\usepackage[T1]{fontenc}
\usepackage{polski}
\usepackage[english]{babel}
\usepackage{graphicx,subcaption,color}
\usepackage{amssymb,amsmath,amsfonts}
\usepackage{cite}
\usepackage[keeplastbox]{flushend}
\usepackage{float}
\usepackage{array}

\usepackage{eso-pic}
\AddToShipoutPicture*{\footnotesize\sffamily\raisebox{0.8cm}{\hspace{1.5cm}\fbox{\parbox{\textwidth}{\copyright~2017
IEEE. Personal use of this material is permitted. Permission from IEEE
must be obtained for all other uses, in any current or future media,
including reprinting/republishing this material for advertising or
promotional purposes, creating new collective works, for resale or
redistribution to servers or lists, or reuse of any copyrighted
component of this work in other works.}}}}

\begin{document}
\newcommand{\corr}[1]{\textcolor{red}{#1}}
\newcommand{\delete}[1]{\st{[#1]}}


\title{STEAM: A Hierarchical Co-Simulation Framework for Superconducting Accelerator Magnet Circuits}

\author{{L. Bortot, B. Auchmann, I. Cortes Garcia, A.M. Fernandez Navarro, M. Maciejewski, M. Mentink,}

{M. Prioli, E. Ravaioli, S. Schöps, and A.P. Verweij}

\thanks{

L. Bortot, M. Prioli, A.M. Fernandez Navarro, M. Mentink and A.P. Verweij are with CERN, Switzerland (e-mail: 	lorenzo.bortot@cern.ch).

B.Auchmann is with CERN, Switzerland, and with Paul Scherrer Institute, 5232 Villigen PSI, Switzerland.

M. Maciejewski is with CERN, Switzerland, and with Institute of Automatic Control, Technical University of Łódź, 18/22 Stefanowskiego St., Poland.

I.C. Garcia and S. Schöps are with Technische Universität Darmstadt, Karolinenpl. 5, 64289 Darmstadt, Germany.

E. Ravaioli is with the Lawrence Berkeley National Laboratory, Berkeley, CA 94720 USA.

The authors I. Cortes Garcia and S. Schöps have been supported by the Excellence Initiative of the German Federal and State Governments and the Graduate School of CE at TU Darmstadt.
}}

\markboth{Tue-Af-Po2.10}
{Shell \MakeLowercase{\textit{et al.}}: Bare Demo of IEEEtran.cls for Journals}


\maketitle
\selectlanguage{English}
\IEEEpeerreviewmaketitle


\begin{abstract}
Simulating the transient effects occurring in superconducting accelerator magnet circuits requires including the mutual electro-thermo-dynamic interaction among the circuit elements, such as power converters, magnets, and protection systems. Nevertheless, the numerical analysis is traditionally done separately for each element in the circuit, leading to possible non-consistent results. We present STEAM, a hierarchical co-simulation framework featuring the waveform relaxation method. The framework simulates a complex system as a composition of simpler, independent models that exchange information. The convergence of the coupling algorithm ensures the consistency of the solution. The modularity of the framework allows integrating models developed with both proprietary and in-house tools. The framework implements a user-customizable hierarchical algorithm to schedule how models participate to the co-simulation, for the purpose of using computational resources efficiently. As a case study, a quench scenario is co-simulated for the inner triplet circuit for the High Luminosity upgrade of the LHC at CERN.
\end{abstract}
	
   
\section{Introduction} \label{Introduction}
Circuits consisting of superconducting accelerator magnets are complex systems that integrate components and technologies belonging to heterogeneous fields of engineering. Each component is coupled with the others, showing mutual interactions. Due to the physical size of the circuit of up to several kilometres, the number of components, and the variety of possible dynamic effects, the simulation of such a system is intrinsically a  multi-physics, multi-scale, and multi-rate problem. In particular, this holds in case of a quench: quench heaters (QH)~\cite{rodriguez2001quench} or the recently developed Coupling-Loss Induced Quench system (CLIQ)~\cite{ravaioli2015cliq} cause transient effects that propagate through the circuit~\cite{ravaioli2012impact}, interacting with the magnets, the power converters and the protection electronics.

\begin{figure}[tb]
  \centering
	\includegraphics[width=8.7cm]{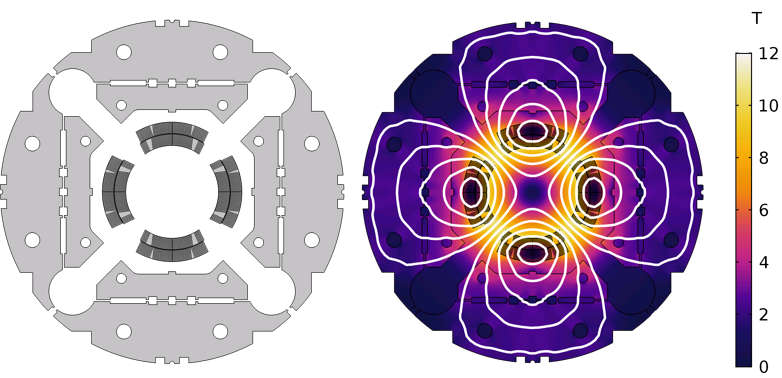}
	\caption{MQXF cross-section (left), and magnetic flux density field at nominal current (right).}
	\label{MQXFGeomAndField}
    \vskip -0.25cm
\end{figure}

Simulations are crucial for understanding the transient phenomena occurring within superconducting accelerator magnet circuits. Numerical methods are widely used in the analysis of both the magnets and the quench protection systems, bringing insights on the quench behaviour and contributing to prevent potentially irreversible consequences. Nevertheless, the currently available high-performance tools cannot capture within one model all the phenomena occurring in an accelerator magnet circuit. Therefore, the system is traditionally decomposed in component-specific models, refined by expert know-how. As a consequence, consistent results are achieved only if all the models' mutual influences are correctly taken into account, with a proper two-way coupling strategy.

These considerations led to the development of STEAM, a co-simulation framework~\cite{henrotte2008efficient,lange2009efficient,zhou2006general,schops2010cosimulation} implemented in Java. 
The key features are a communication bus and a user-customizable hierarchical algorithm. The former exchanges information between multiple models, the latter schedules how the models participate to the co-simulation, solving them only when needed for the accuracy of solution. 
The coupling of the models occurs via a dedicated algorithm implementing the waveform relaxation method~\cite{white1987waveform}. The convergence of the coupling algorithm ensures the consistency of the solution. The framework architecture is expandable and can support both proprietary and in-house tools.

In this paper we introduce the core algorithms and the architecture of the framework. Then a case-study illustrates the decomposition of the system, the choice of the solvers, and the hierarchical organization of the models. The case-study assumes a quench occurring in one of the $\mathrm{Nb_{3}Sn}$ quadrupole magnets (MQXF)~\cite{ferracin2016development} (see Fig.~\ref{MQXFGeomAndField}) belonging to the future inner triplet circuit for the High Luminosity upgrade of the LHC at CERN~\cite{hilumi2014hl}. The relevance of achieving consistent simulations is discussed in the results section.

\section{Framework Implementation} \label{FrameworkImplementation}
The hierarchical co-simulation approach turns the analysis of a circuit of accelerator magnets into a coupled problem~\cite{park1983partitioned}. The complexity of the original system is represented through a composition of simpler models. The waveform relaxation method~\cite{schops2010cosimulation} is applied to resolve the mutual dependencies between the models, with the following algorithm. i) The overall simulation time is divided into windows. ii) Within a  window, the models are solved and the relevant waveforms are exchanged, following a Gauss-Seidel scheme~\cite{burrage1995parallel}. iii) The previous operation is repeated for the same window until the waveforms belonging to two consecutive iterations differ by less than a prescribed tolerance, then the algorithm moves to the next window. The convergence of the algorithm for every window ensures the consistency of the overall solution~\cite{bartel2013dynamic}.

\subsection{Architecture} \label{Architecture}
\begin{figure}[tb]
  \centering
	\includegraphics[width=8.0cm]{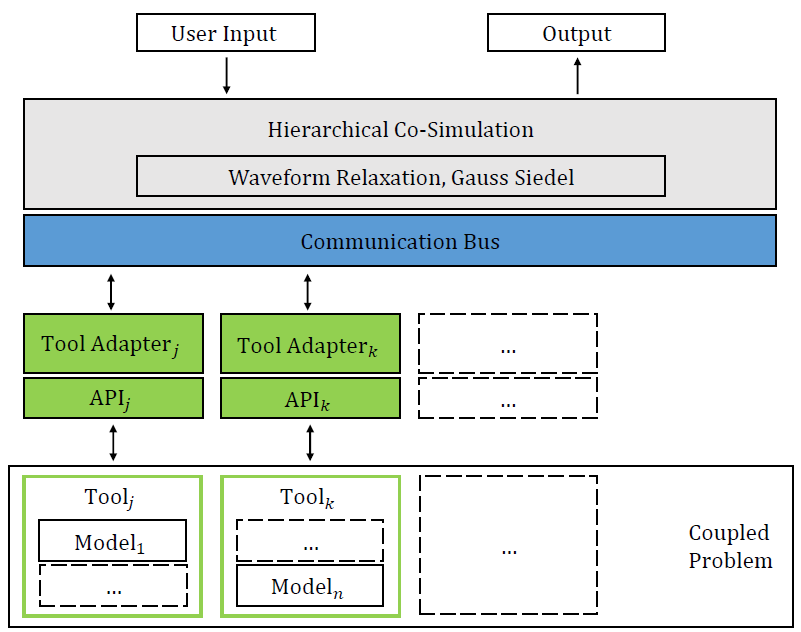}
	\caption{Framework architecture. A generic coupled problem is decomposed in $n$ models developed with $k$ different tools.}
	\label{HierarchicalCosimulationStructure}
\end{figure}

The STEAM framework is developed on a three-layer, scalable and expandable structure (see Fig.~\ref{HierarchicalCosimulationStructure})~\cite{Maciejewski2017STEAMarchitecture}:
\newline
1) The top layer contains the hierarchical co-simulation algorithm implementing the waveform relaxation method. The functionalities of the layer are to manage the  execution of the models over the simulation time windows, to check the waveform relaxation convergence and to provide the necessary input/output interfaces. 
\newline
2) The middle layer exchanges information between the models that participate to the co-simulation. The layer is composed by a communication bus which expects a specific communication protocol. The bus can handle both time- and space-dependent signals. For the latter ones, the $\mathrm{MpCCI}$~\cite{ahrem2000mpcci} mesh-based interpolation tool is in use \cite{Maciejewski2017MeshInterpolation}.
\newline
3) The bottom layer implements a modular structure, composed by blocks called \textit{tool adapters}. Each adapter exchanges signals between the communication bus and the models belonging to a tool via a suitable Application Programming Interface (API), which is tool-dependent. This ensures that different models developed within the same tool rely on the same tool adapter. The architecture is scalable and expandable: new simulation tools can be interfaced with the framework by developing the dedicated tool adapters.

\begin{figure}[h]
  \centering
	\includegraphics[width=8.0cm]{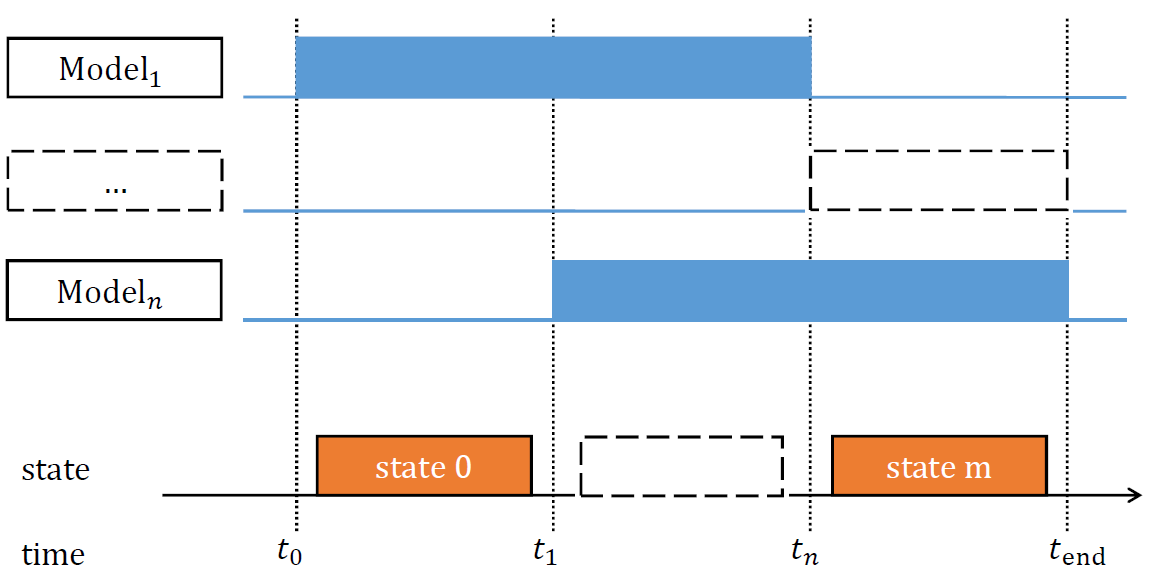}
	\caption{Hierarchical organization of the $n$ models representing a time-dependent coupled problem, in $m$ states.}
	\label{HierarchicalStructure}
\end{figure}

\begin{figure*}[tb]
\centering
\includegraphics[width=16.0cm]{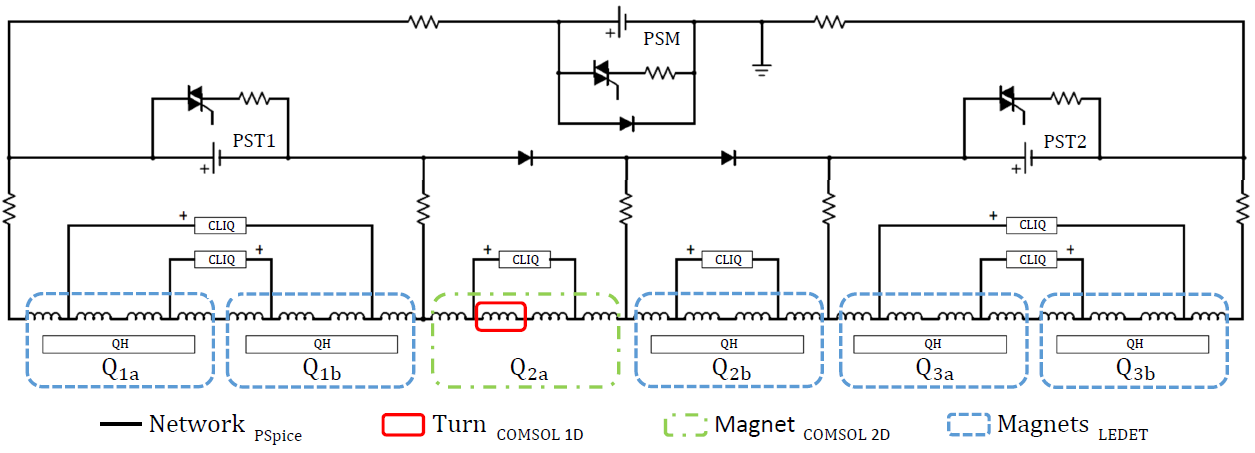}
\caption{Inner triplet circuit. the system decomposition and the choice of the tools are highlighted.}
\label{TripletCircuitDecomposition}
\vskip -0.25cm
\end{figure*}

\subsection{Hierarchy} \label{Hierarchy}
The transient phenomena characterizing a circuit of accelerator magnets might occur at different times, with different durations. If these phenomena are distributed among separated multiphysical models, then possibly not all the models are necessary during the full co-simulation timespan. This observation justifies the introduction of a hierarchical state-machine algorithm for the models' management. Referring to  Fig.~\ref{HierarchicalStructure}, the user defines both the states and the transitions of the system. The simulation time is separated into states, and the subset of active models are determined for each state. A transition can be defined by a fixed time, or a conditional expression dependent on the waveforms exchanged among the models. The benefit is twofold: the state machine input explicitly determines the causality relations existing between the models, and the overall computational cost is reduced.

\section{Case Study} \label{CaseStudy}
As an example, STEAM is used for analysing a quench scenario of the future inner triplet circuit for the High Luminosity LHC. The circuit powering scheme requires three power supply units (see Fig.~\ref{TripletCircuitDecomposition}): a main one ($\mathrm{PS{M}}$) on the outer current loop and two trim units ($\mathrm{PS{T1}}$, $\mathrm{PS{T2}}$) located in nested current loops. Moreover, the system contains nonlinear bypass components such as flywheel diodes and crowbars, and six newly designed $\mathrm{Nb_{3}Sn}$ magnets $\mathrm{Q_{1a}}$, $\mathrm{Q_{1b}}$, $\mathrm{Q_{2a}}$, $\mathrm{Q_{2b}}$, $\mathrm{Q_{3a}}$, $\mathrm{Q_{3b}}$. The magnets are protected by a combination of QHs and CLIQ units, with the latter ones connected over single and multiple magnets~\cite{ravaioli2017quench}. Furthermore, the circuit protection strategy requires the simultaneous intervention of the protection systems for all the magnets, in case of a quench. As a consequence, the mutual dependencies between the system components justify the co-simulation approach for the numerical analysis of the scenario.

In the case study, the current in the circuit is ramped-up to  nominal conditions, with $\mathrm{I_{PSM}=16.47\ kA}$, and $\mathrm{I_{PST1}=I_{PST2}=-2\ kA}$. 
Subsequently, one of the high-field turns of $\mathrm{Q_{2a}}$ is assumed to quench. The longitudinal propagation of the quench within the conductor and the steadily increasing temperature result in a growing resistive voltage. At a certain moment the quench is detected: the power converters are then switched off, the crowbars are activated and the quench protection systems of the magnets are triggered. In order to give generality to the method, a malfunction in the QHs of $\mathrm{Q_{2a}}$ is assumed. 

\subsection{System decomposition} \label{SystemDecomposition}
\begin{figure}[tb]
  \centering	\includegraphics[width=7.5cm]{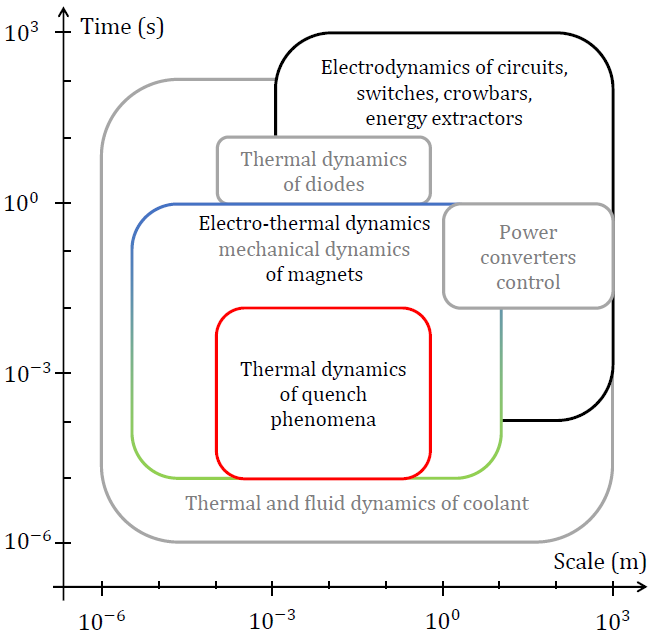}
	\caption{Transient phenomena occurring in a generic superconducting accelerator circuit, represented in a time-scale reference frame.}
	\label{TripletTimeScaleDecomposition}
    \vskip -0.25cm
\end{figure}

The inner triplet system decomposition is tailored to capture only the transient phenomena which are relevant for the given quench scenario (see Fig.~\ref{TripletTimeScaleDecomposition}). In particular, the study focuses on simulating the initial quench propagation, the magnetothermal dynamics of the magnets and the electrical behaviour of the network. Nevertheless, the system decomposition can be refined with more dedicated models, to include other devices and physical phenomena. As an example, one can include the thermal behaviour of the bypass diodes, the action of the digital controllers of the power converters~\cite{maciejewski2017application}, or the mechanical response of the magnets~\cite{Maciejewski2017MeshInterpolation}. 

The system is partitioned in three sub-units (see Fig.~\ref{TripletCircuitDecomposition}): the magnet turn  where the initial quench occurs, the magnets and the remaining network.
1) {The quenching turn is simulated with a dedicated 1D adiabatic model, due to the local nature of the quench initiation and propagation. Since the heat propagation between adjacent turns is neglected, the model provides conservative results;}
2) {The quadrupole magnets are represented by individual models simulating the magnetothermal transient induced by the action of the protection systems;}
3) {The network (see Fig.~\ref{TripletCircuitDecomposition})  provides the electrical coupling between all the components in the circuit. Each component is included in the network using an equivalent lumped-parameters representation. An additional a Java code simulates the quench detection signals associated to each magnet (Sec.~\ref{Result}).}

\subsection{Choice of tools} \label{ToolsChoice}
As a consequence of the system decomposition, each model covers only a subset of transient phenomena. At this point, the most suitable simulation tools are determined for each subset. 1) {A SPICE~\cite{vladimirescu1994spice} tool is used to calculate the currents and voltages of the inner triplet equivalent network. The proprietary distribution $\text{PSpice}^{\circledR}$~\cite{orcad1978ppspice} has been used, although the freeware version $\text{LTspice}$~\cite{engelhardt2011ltspice} is also supported;} 2) {The finite-element (FE) proprietary tool $\text{COMSOL}^{\circledR}$~\cite{comsol2005comsol} is used to calculate the quench initiation and propagation, and the consequent resistive voltage;} 3) {${\mathrm{Q_{2a}}}$ is represented with a $\text{COMSOL}^{\circledR}$ magnetothermal model, which implementation details are described in ~\cite{Bortot2017fem}. The {FEM} method is chosen for providing a detailed analysis of the quenching magnet. The 2D representation is justified due the large length/diameter ratio of the coils, and the uniform energy deposition along the coils by means of CLIQ; 4) The other magnets are modelled using LEDET~\cite{ravaioli2016lumped,ravaioli2015cliq}. In particular, the magnetothermal formulation is implemented using equivalent networks of lumped elements which solve faster in comparison to the FE model.

\subsection{Hierarchical organization of the models} \label{Hierarchization}
The models are organized in a hierarchical structure, reflecting the different system's states during the co-simulation (see Fig.~\ref{HierarchicalCoSimulationUnitsLogicStates}).
At $t_{\mathrm{0}}$ the circuit is powered: only the network solver is required to ramp up the circuit to nominal operation conditions. At $t_{\mathrm{quench}}$ the quench is introduced and the 1D model is activated. The resistive voltage grows until it is detected, at $t_{\mathrm{discharge}}$. At this point the 1D model is disabled. 
The dynamics of the circuit is determined by the action of the quench protection systems on the quadrupole magnets. For this reason, the dedicated 2D MQXF models are enabled and kept active until $t_{\mathrm{end}}$, when the energy in the circuit is completely discharged. 
\begin{figure}[tb]
  \centering
	\includegraphics[width=8.0cm]{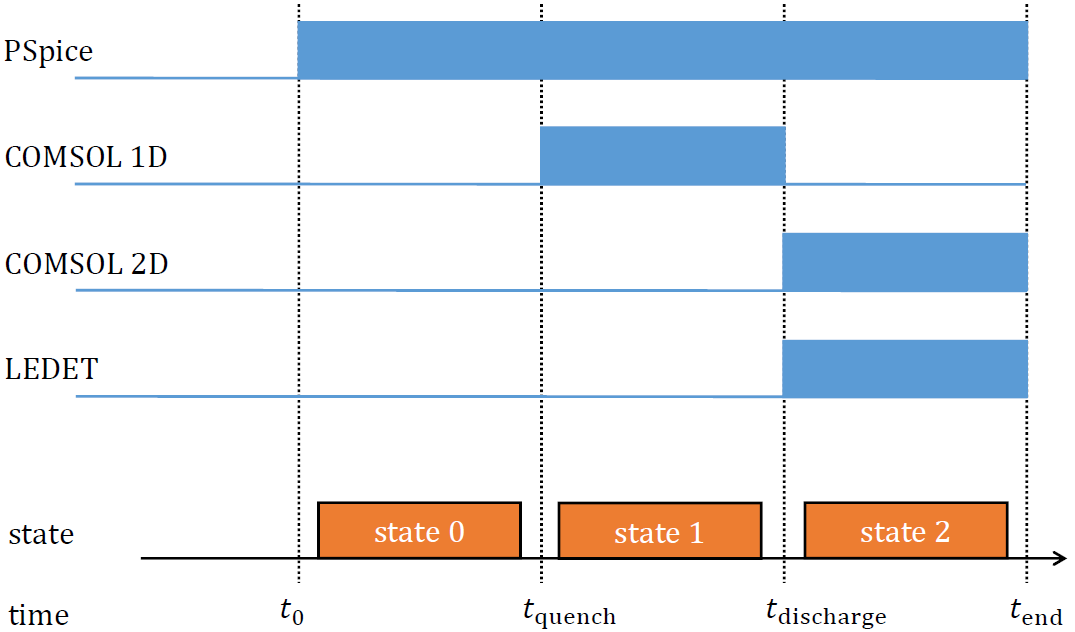}
	\caption{Hierarchical organization of the models, as function of the simulation time.}
	\label{HierarchicalCoSimulationUnitsLogicStates}
    \vskip -0.25cm
\end{figure}

\section{Results} \label{Result}
\begin{figure}[tb]
  \centering
	\includegraphics[width=8.5cm]{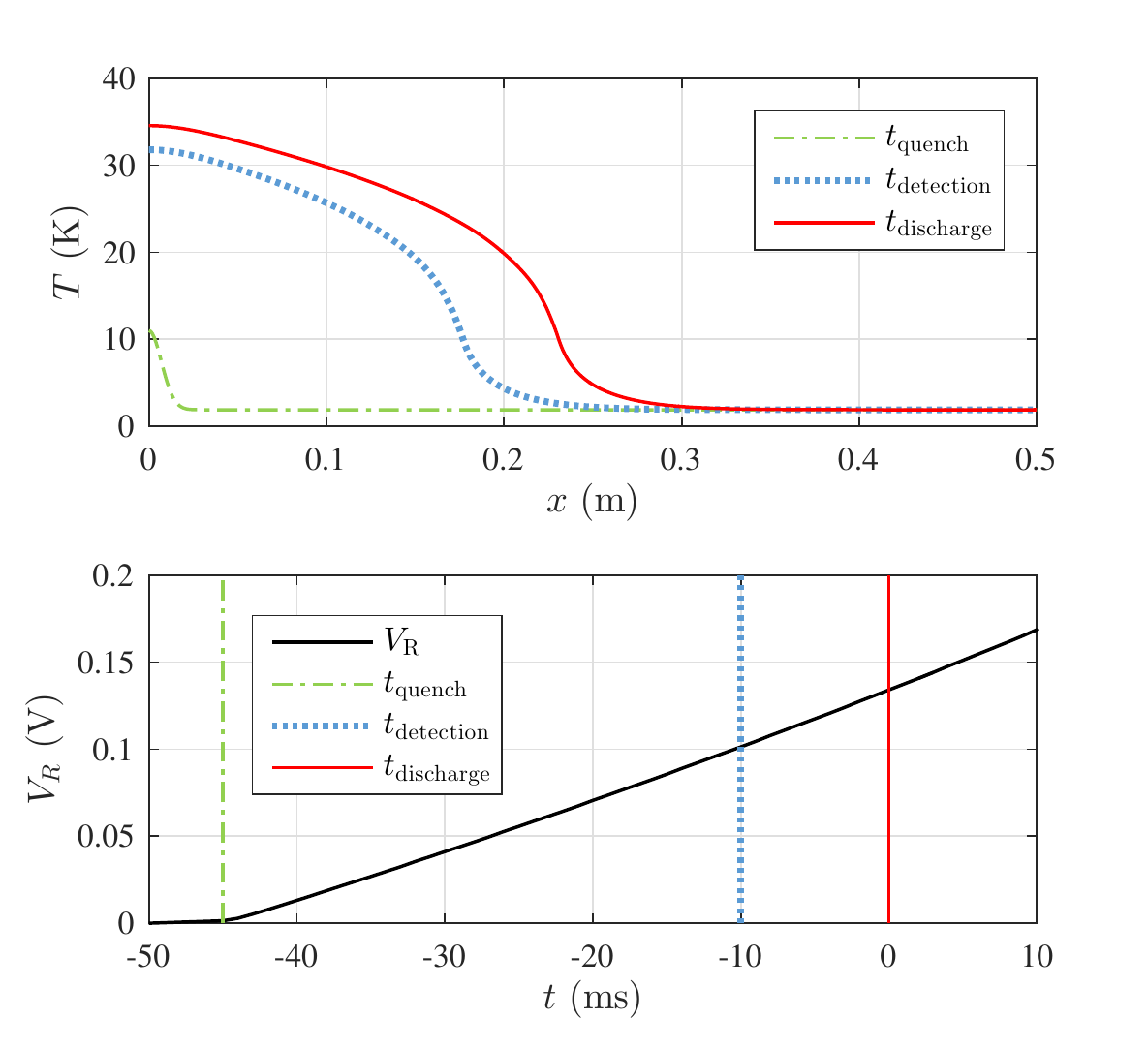}
	\caption{Top: temperature profile along the quenching turn, during the quench propagation. Bottom: Evolution of the associated resistive voltage.}
	\label{V_quench}
    \vskip -0.25cm
\end{figure}
\begin{figure}[tb]
\centering
\captionsetup[subfigure]{{skip=-0.5cm}}
\captionsetup[subfigure]{position=top,singlelinecheck=off,justification=raggedright}
    \begin{subfigure}[tb]{0.5\textwidth}
   		\caption{}
   		\label{fig:No1a}
        \centering
   		\includegraphics[width=8.0cm]{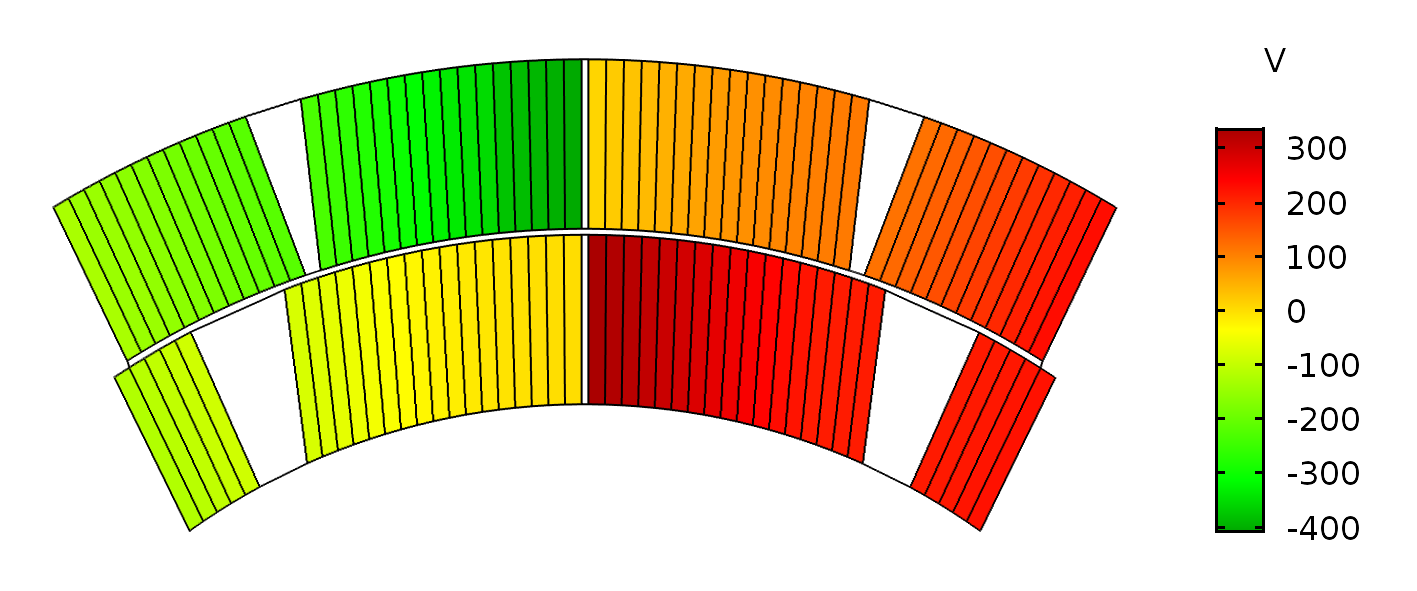}
	\end{subfigure}
	\begin{subfigure}[tb]{0.5\textwidth}
        \caption{}
        \label{fig:No2a}
        \centering
   		\includegraphics[width=8.0cm]{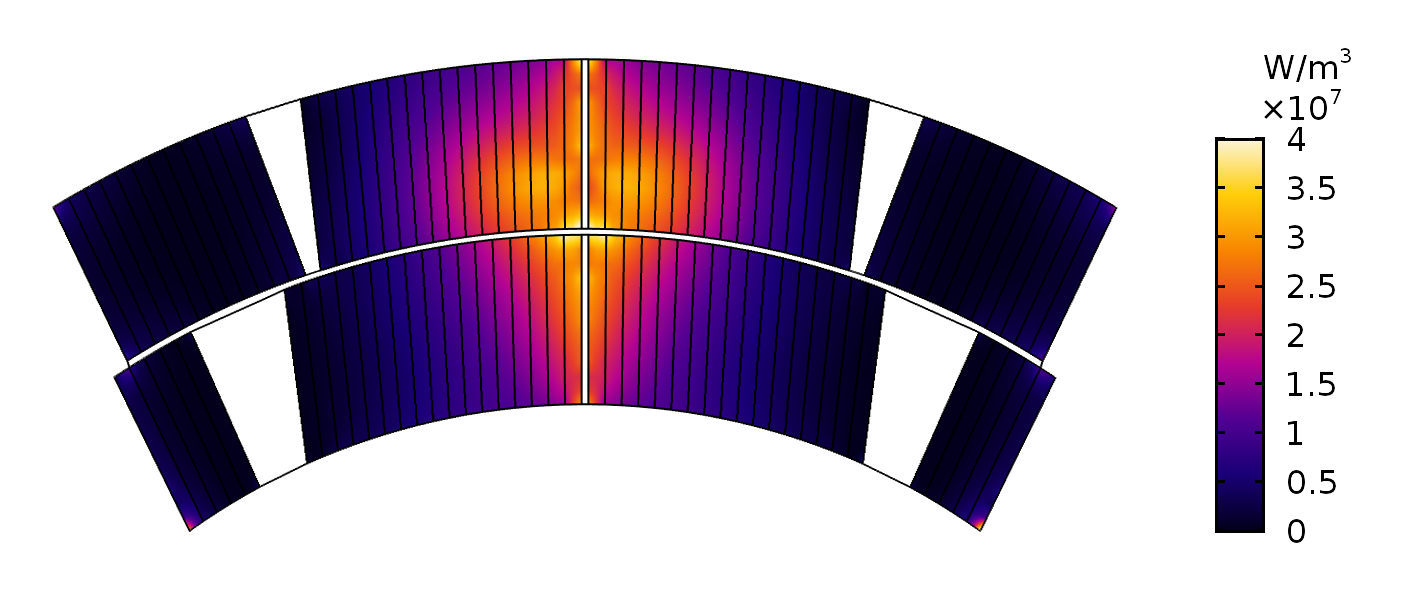}
	\end{subfigure}
	\begin{subfigure}[tb]{0.5\textwidth}
   		\caption{}
   		\label{fig:No4a}
        \centering
   		\includegraphics[width=8.0cm]{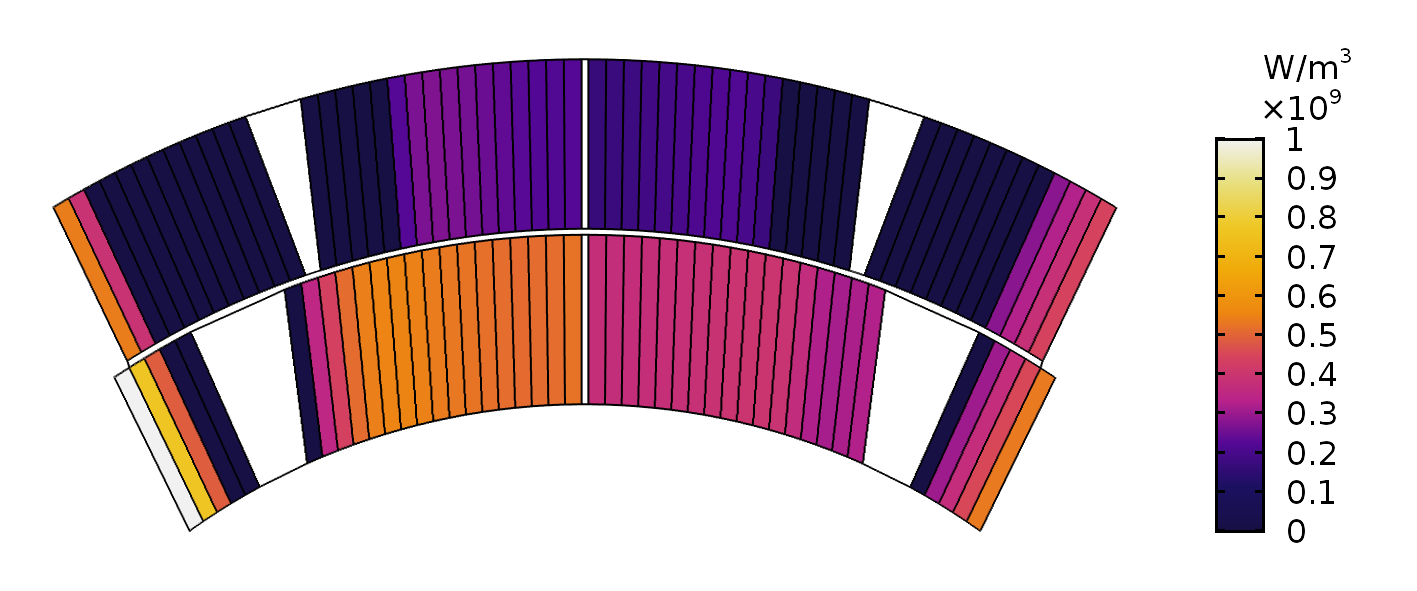}
	\end{subfigure}
   \begin{subfigure}[tb]{0.5\textwidth}
   		\caption{}
   		\label{fig:No5a}
        \centering
   		\includegraphics[width=8.0cm]{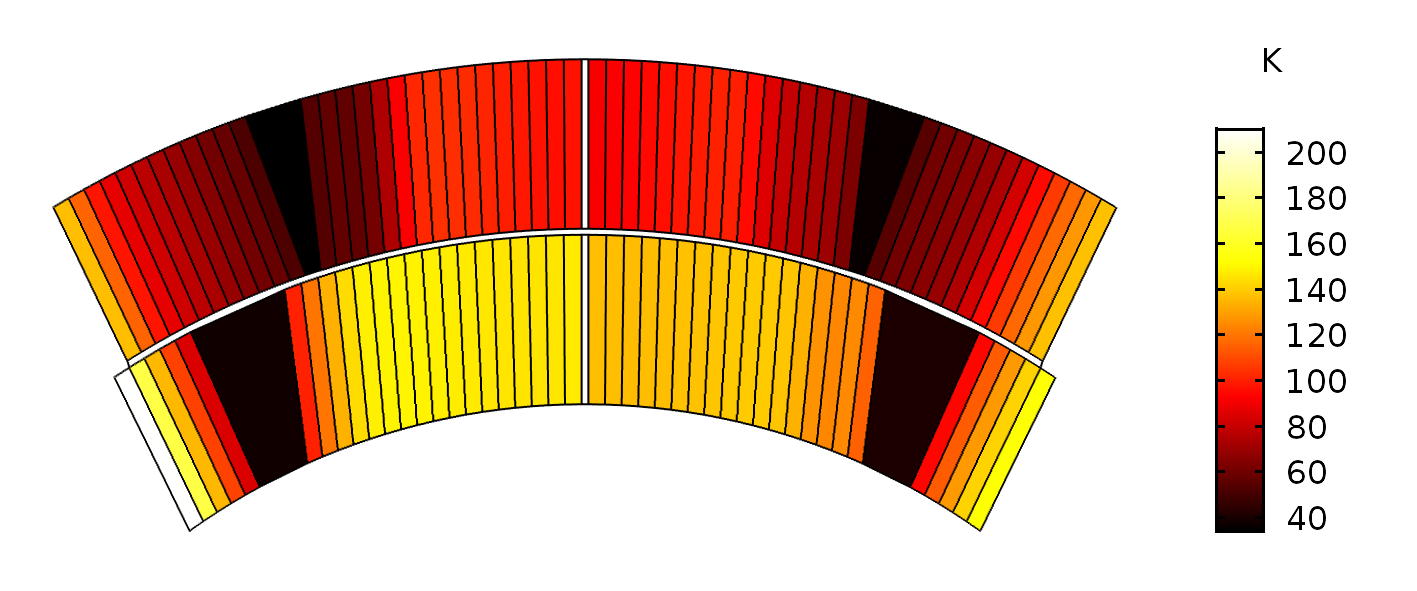}
\end{subfigure}
\caption{ (a) Voltage-to-ground distribution, 5 ms after $\mathrm{t_{discharge}}$.
(b) Inter-filament and inter-strand coupling losses,  5 ms after $\mathrm{t_{discharge}}$. 
(c) Ohmic losses,  25 ms after $\mathrm{t_{discharge}}$. 
(d) Temperature distribution, 500 ms after $\mathrm{t_{discharge}}$.
}
\label{MQXF_EMTH_Transient}
\vskip -0.25cm
\end{figure}

The results focus on the dissipation of the energy stored in the circuit after $t_{\mathrm{quench}}$. Firstly, the magnetothermal transient occurring in $\mathrm{Q_{2a}}$ is discussed in detail. Next, two magnet-network coupling strategies are applied, the former based on an equivalent thermal representation and the latter on the field-circuit coupling technique. Lastly, the results given by the two coupling strategies are compared.

The magnet $\mathrm{Q_{2a}}$ is assumed to quench at $t_\mathrm{{quench}}$ (Fig.~\ref{V_quench}, top),  developing a resistive voltage contribution $V_\mathrm{R}$ (Fig.~\ref{V_quench}, bottom). The normal zone propagates along the turn until the quench detection threshold ($V_\mathrm{R}>100\mathrm{\ mV}$) is reached, at $t_\mathrm{{detection}}$. Subsequently, a validation criterion is applied to $V_\mathrm{R}$, requiring the resistive voltage signal to exceed a threshold of $\mathrm{100\ mV}$ for $\mathrm{10\ ms}$. Once the quench is validated, the protection systems are activated at $t_\mathrm{{discharge}}$.

The  1000~V~/~40~mF CLIQ unit is discharged in parallel to the  $\mathrm{Q_{2a}}$ coil poles. This induces a voltage redistribution (Fig.~\ref{MQXF_EMTH_Transient}a), turning into an imbalance in the poles currents which oscillate due to the resonance between the inductive and capacitive behaviour of the magnet and CLIQ unit, respectively. 
The oscillation of the magnetic field determines the dissipation within the coil of the energy stored in the CLIQ unit, as inter-filament and inter-strand (Fig.~\ref{MQXF_EMTH_Transient}b) coupling currents~\cite{wilson1983superconducting,verweij1995electrodynamics}.
The energy deposition determines the quench of a bigger volume of the coil, which becomes resistive and contributes both to the discharge of the magnet through the dissipation of  Joule losses (Fig.~\ref{MQXF_EMTH_Transient}c), and to a temperature increase in the coil (Fig.~\ref{MQXF_EMTH_Transient}d). The  computational time of the co-simulation is about 5h, on a standard workstation.

\begin{figure}[tb]
\centering
\captionsetup[subfigure]{{skip=-0.4cm}}
\captionsetup[subfigure]{position=top,singlelinecheck=off,justification=raggedright}

    \begin{subfigure}[tb]{0.5\textwidth}
   		\caption{}
   		\label{fig:No1b}
        \centering
   		\includegraphics[width=8.0cm]{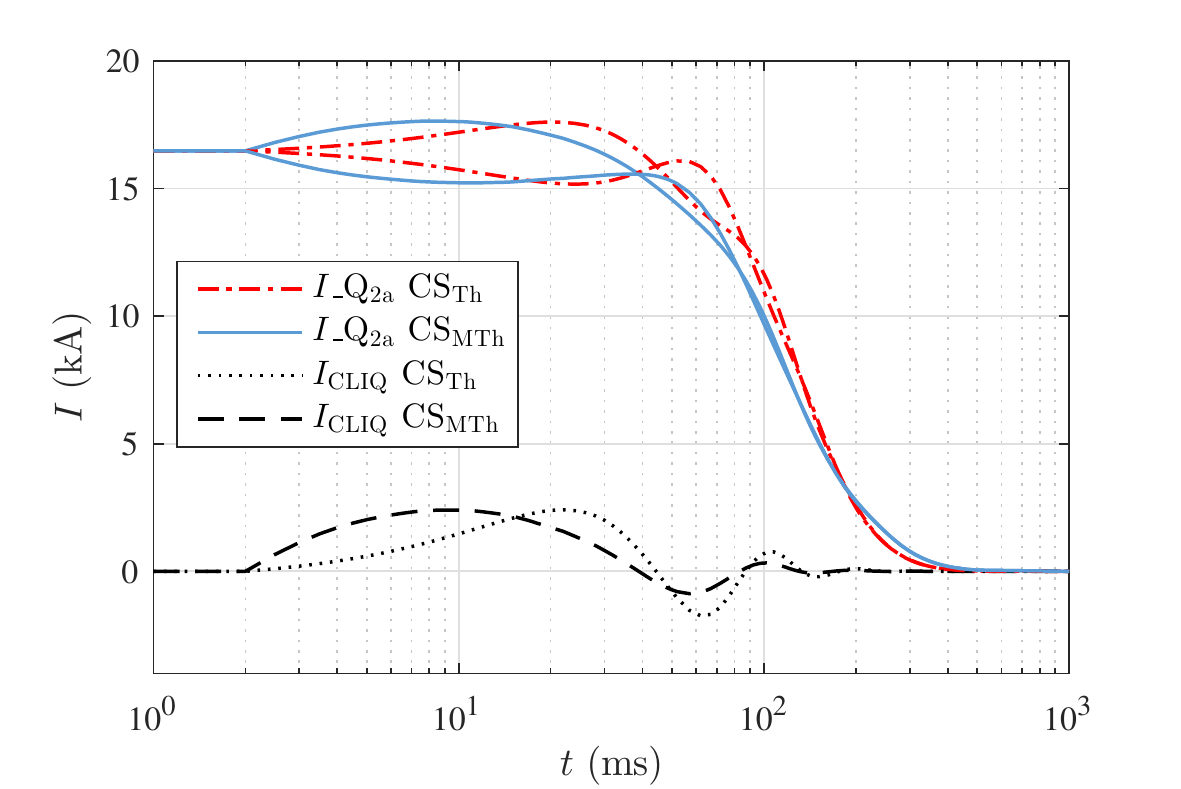}
	\end{subfigure}
	\begin{subfigure}[tb]{0.5\textwidth}
        \caption{}
        \label{fig:No2b}
        \centering
   		\includegraphics[width=8.0cm]{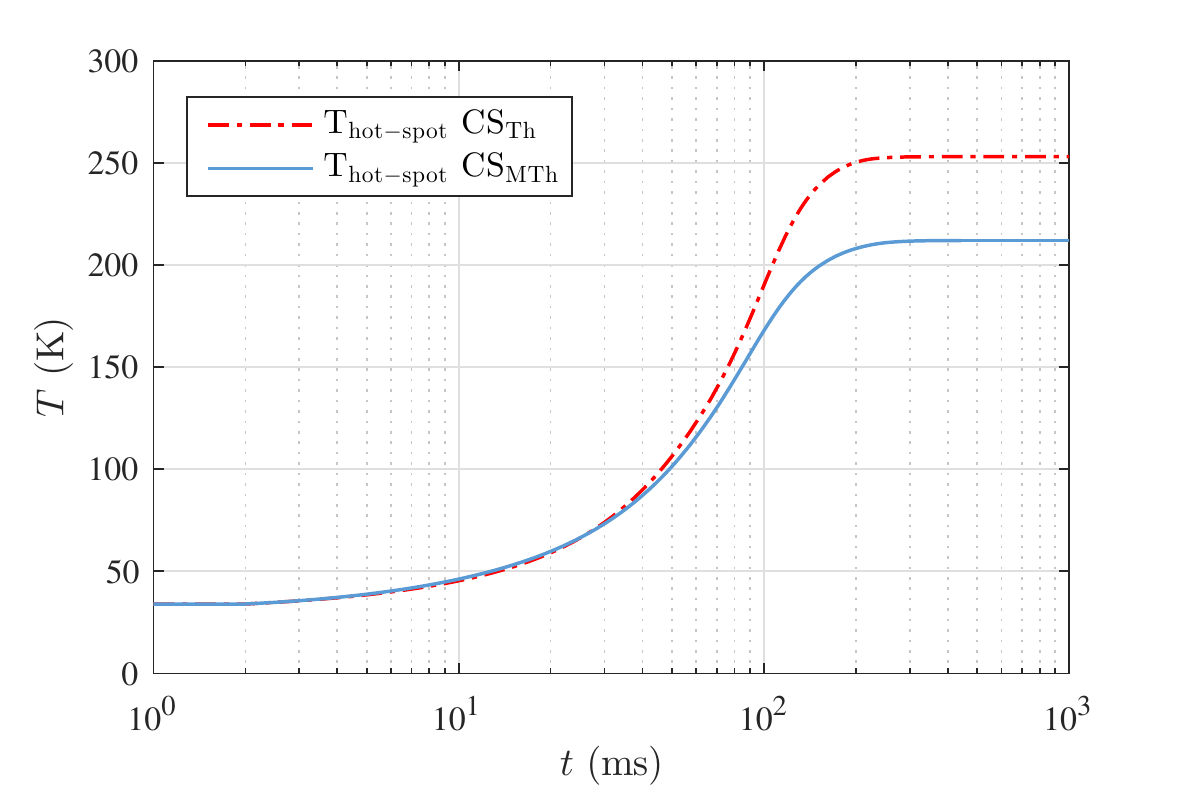}
	\end{subfigure}

	\begin{subfigure}[tb]{0.5\textwidth}
   		\caption{}
   		\label{fig:No4b}
        \centering
   		\includegraphics[width=8.0cm]{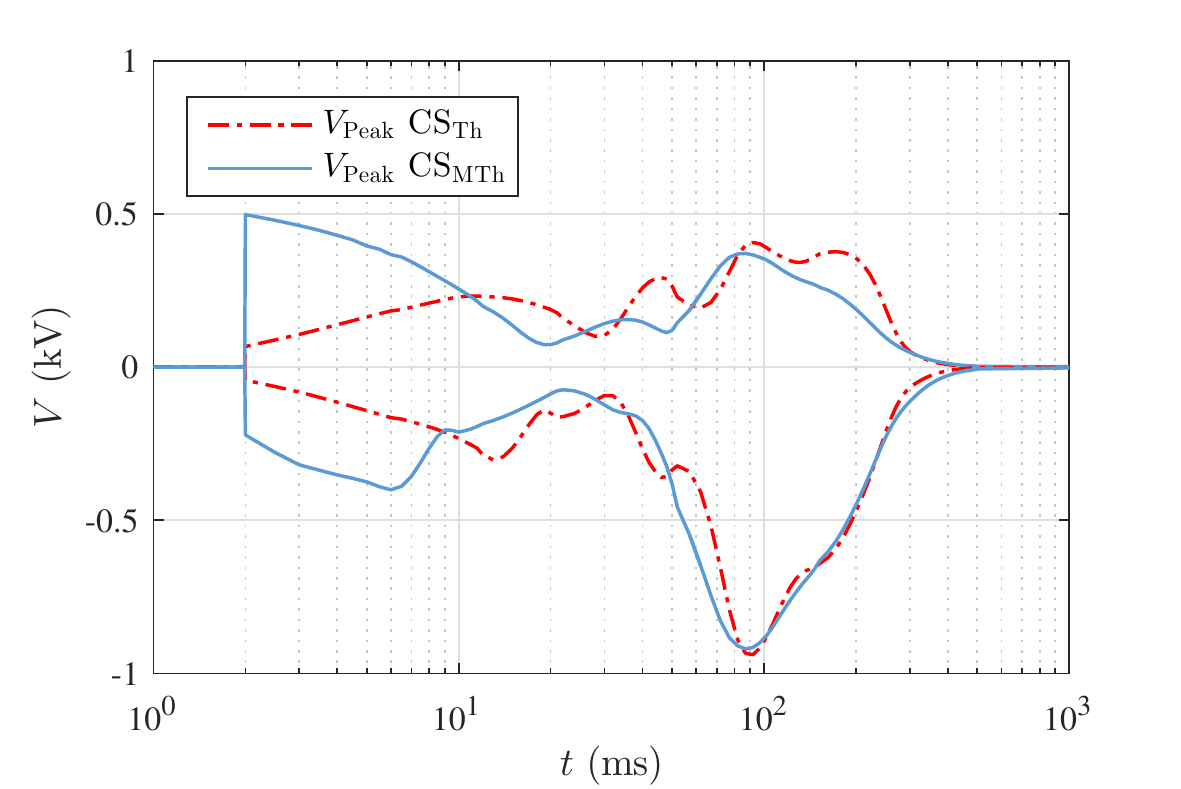}
	\end{subfigure}
\caption{ $\mathrm{Q_{2a}}$ magnet. (a) Currents in the  coil and in the CLIQ unit;
(b) Hot-spot temperature;
(c) Maximum and minimum voltage to ground.
}
\label{CoSimAnalysis}
\vskip -0.25cm
\end{figure}

The magnets' internal dynamics imposes the equivalent impedances seen by the network. The network, in turn, determines the currents driving the magnets' dynamics. Hence, a reliable quench protection simulation requires a consistent two-way coupling between the magnets and the circuit. To prove this, a comparison is provided. In the co-simulation $\mathrm{{CS}_{Th}}$, each magnet is represented in the network as a linear inductor in series with a time varying resistor. Such a simplification takes into account the thermal response of the magnet but neglects both the saturation and the dynamic phenomena in the superconducting coils. In the co-simulation $\mathrm{{CS}_{MTh}}$, the dynamics of each magnet is consistently represented using the field-circuit coupling technique~\cite{schops2010cosimulation,garcia2017optimized,bortot2017consistent}.

The currents for $\mathrm{Q_{2a}}$ and the related CLIQ unit are shown in Fig.~\ref{CoSimAnalysis}a), for both the co-simulations. The coil dynamics greatly reduces the magnet's differential inductance, determining steeper initial derivatives in the currents of $\mathrm{{CS}_{MTh}}$. This, in turn, causes a higher deposition of dynamic losses, a more homogeneous spread of the quench, a faster current decay and a lower Joule integral (MIITs). This is reflected both in Fig.~\ref{CoSimAnalysis}b) where the hot-spot temperature decrease by $\mathrm{42\ K}$, and in Fig.~\ref{CoSimAnalysis}c), where the peak voltage to ground decrease by $\mathrm{15\ V}$. A summary of the comparison is provided in Tab.~\ref{TablePeakValues}, showing an overall reduction in the estimation of the magnet stress parameters.

\begin{table}[!h]
\setlength{\extrarowheight}{3pt}
\centering
\caption{Comparison of results}
\begin{tabular}{lccccc}\hline\hline
{Quantity} &  {Unit} & {$\mathrm{CS_{Th}}$} & {$\mathrm{CS_{MTh}}$} & {$\Delta$} & {$\Delta\ \%$} \\ \hline
$\mathrm{MIITs}$ & $\mathrm{MA^{2}s}$ & 27 & 24 & -3 & -9 \\
$\mathrm{T_{Hotspot}}$ & $\mathrm{K}$ & 253 & 211 & -42 & -16  \\
$\mathrm{V_{PeakGnd}}$ & $\mathrm{V}$ & 935 & 920 & -15 & -2 \\ \hline \hline
\end{tabular}
\label{TablePeakValues}
\end{table}

\section{Conclusions and outlook} \label{ConclusionAndOutlook}
The role of simulations is twofold. On the one hand, numerical methods provide support to the design of both circuit components and quench protections systems. On the other hand, they bring insights to the transient phenomena occurring in superconducting accelerator magnet circuits, even for quantities that cannot be measured. This holds true for the high-performance, low-margin inner triplet circuit for the LHC High Luminosity upgrade. The design requires accurate simulations, due to the mutual electro-thermal coupling occurring among the magnets, the quench protection systems and the rest of the network. Simulating such a complex system in a single tool with high accuracy is currently not feasible. 

For this reason, we propose STEAM, a Java-developed framework which allows to decompose the original system into simpler, independent models solved consistently, as a coupled problem. The hierarchical algorithm ensures an efficient use of computational resources, enforcing the models to be solved only for the simulation time span where they contribute to the accuracy of the solution.
 The framework architecture is scalable and expandable, giving the possibility to add further proprietary and in-house solvers in the future. A quench protection scenario in the inner triplet circuit is used as a case study, to illustrate the co-simulation approach. Results remark the importance of taking into account the mutual influences between the sub-units in the circuit, in a consistent way. The analysis of the magnets' dynamics is improved, leading to less conservative results: welcomed margins are highlighted in the baseline design of the high-performance MQXF magnets.

STEAM is actively supporting the simulation needs of the most demanding accelerator projects~\cite{valette2017effect,fernandez201711T,prioli2017strategies,mentink2017blockCoil}. The framework is ready to be extended to new modules and tools, to cover more simulation needs: examples are the QHs dynamics, the mechanical response of the magnet coils, the 3D quench propagation, and the fluid dynamics of the coolant.

\section{Acknowledgement} \label{Acknowledgement} 
The authors would like to thank K. Kr{\'o}l, and J.C. Garnier from CERN for the continuous help provided in developing the STEAM Framework, and S. Yammine from CERN for the support in the development of the inner triplet PSpice model.

\clearpage
\newpage


\end{document}